# A MODEL FOR THE RADIO EMISSION FROM SNR 1987A

P. Duffy

*Max-Planck-Institut für Kernphysik, W-69029 Heidelberg, Germany.*
*I: duffy@peter.mpi-hd.mpg.de*

Lewis Ball

*Research Centre for Theoretical Astrophysics,*
*University of Sydney, N.S.W. 2006, Australia.*
*I: ball@physics.su.oz.au*

J. G. Kirk

*Max-Planck-Institut für Kernphysik, W-69029 Heidelberg, Germany.*
*I: kirk@kirk0.mpi-hd.mpg.de*

## ABSTRACT

The observations of radio emission from SNR 1987A can be accounted for on the basis of diffusive shock acceleration of electrons by the supernova blast wave. However, with this interpretation the observed spectral index implies that the compression ratio of the gas subshock is roughly 2.7 rather than the value of 4 expected of a strong shock front. We propose that in SNR 1987A, ions also undergo diffusive acceleration at the shock, a process that is likely to be rapid. Unlike the electron population, the accelerated ions can have an important effect on the gas dynamics. We calculate this coupled gas and energetic particle dynamics on the basis of the two-fluid model, in which the accelerated ions provide an additional component to the total pressure acting on the fluid. By accelerating and possibly heating the upstream plasma, the initially strong shock is modified and a weaker subshock with an upstream precursor results. The electrons behave as test particles. They are accelerated at the evolving subshock, escape downstream, and emit synchrotron radiation in the swept up magnetic field. Two models are considered for the surroundings of the progenitor: that of a freely expanding wind of number density $n \propto r^{-2}$, and that of a wind confined by a shell of denser material, creating a stagnation zone of roughly constant density beyond the standing shock which terminates the free wind. We model the observed radio light curves and the relatively steep spectrum of SNR 1987A using similar values for the ion acceleration parameters to those used in models of cosmic ray acceleration in older SNRs which can also contain high Mach number shocks, and find a good fit for the case in which the termination shock is located at about $2 \times 10^{15}$ m from the progenitor.

*Subject headings:* acceleration of particles, shock waves, supernovae : general, supernovae : individual : SN1987A, supernova remnants

## 1. Introduction

The intensity of the radio emission from the remnant of the supernova SN1987A has increased steadily ever since it was detected in July 1990 (Staveley-Smith et al. 1992). This behaviour is unlike that observed from other radio supernovae – which generally exhibit a rapid rise of radio emission followed by a steady decline (Weiler et al. 1986) – and has stimulated new ideas on the origin of the radiation and the interaction of this supernova with its surroundings (Chevalier 1992a, and Ball & Kirk 1992a – hereafter BK). The latter paper proposed the diffusive acceleration of electrons and their subsequent synchrotron radiation as the mechanism responsible for the radio emission, and modelled the light curve up to day 1800 by assuming electron injection takes place in two distinct components or clumps. This is consistent with later observations (Staveley-Smith et al. 1993) showing that at about day 2100 the image of the remnant consisted of two hot-spots, or clumps of emission, superimposed on a diffuse, more or less spherical background. However, the model suggests that the picture of the supernova shock as a strong, unmodified shock front in a gas of adiabatic index 5/3, is overly simplistic. The theory of diffusive shock acceleration makes a definite prediction of the spectral index given the compression ratio of a simple unmodified shock (see Jones and Ellison 1992 and Blandford and Eichler 1987 for reviews). On the basis of the spectral index of the radio emission, the model fits of BK suggest that the electron acceleration takes place at a shock of compression ratio $\sim 2.7$.

In a recent paper (Kirk, Duffy & Ball 1994) we extended this model by proposing that the shock front is not a simple discontinuity travelling out into the undisturbed gas of the progenitor's stellar wind, but has a structure which is modified by a substantial population of accelerated ions. These particles build up a precursor to the discontinuity (or 'subshock') in the gas flow and lead to a reduction of its compression ratio. Electrons have a mean free path which is too short to permit diffusion from the subshock into the precursor. They are accelerated by the subshock which, having a lower compression ratio than an unmodified shock (and than the overall precursor-subshock structure), leads to a relatively steep electron spectrum (Bell 1987 and Ellison & Reynolds 1991) in agreement with that inferred from the synchrotron spectrum. If our model is correct, it has important implications not only for the future emission of SNR 1987A, but also for the theory of other radio supernovae (RSNe) as well as for the theory of the origin of cosmic rays (in the energy range below $10^{15}$ eV). The purpose of this paper is to present a detailed analysis of the model and compare its predictions with the most recent observations of radio emission from SNR 1987A.

The properties of the medium into which the shock front propagates have a strong influence on the expected light curve. The model of BK assumes a freely-expanding stellar wind around the supernova, which implies a number density upstream of the shock front proportional to $r^{-2}$ (where $r$ is the radius), and (in the absence of reconnection or dynamo action) a magnetic field $B \propto r^{-1}$. However, there is substantial evidence that the progenitor of SN1987A was, until roughly $10^4$ years ago, surrounded by a dense, slow wind typical of a red giant star (for a review see McCray 1993). The transition to the blue-giant phase, in which the star found itself upon explosion, was presumably accompanied by an increase of the wind velocity at the stellar surface from the $10 \, \text{km s}^{-1}$ appropriate for a red-giant wind, to about $500 \, \text{km s}^{-1}$ (Chevalier & Fransson 1987). In such a configuration the blue-giant wind is expected to inflate a bubble inside the relatively dense red-giant wind (Luo & McCray 1991; Blondin & Lundquist 1993). Assuming spherical symmetry, one would expect the environment of the progenitor to have had an inner zone of freely-expanding plasma (the blue giant wind) with number density $n \propto r^{-2}$ and field $B \propto r^{-1}$ extending out to a shock front at $r = r_\text{w}$ (known as the termination shock), surrounded by a stagnation zone with roughly constant density and a magnetic field $B \propto r$ (the shocked blue-giant wind). The outer boundary of the stagnation zone is then formed by a contact discontinuity separating the shocked blue-giant wind from a dense shell of material swept up from the red-giant wind.

The environment around SN1987A shows a marked departure from spherical symmetry. Optical observations show a ring of material of radius $6.3 \times 10^{15}$ m (Jakobsen et al. 1991; Panagia et al. 1991) and a bipolar nebula structure extending some $2.5 \times 10^{16}$ m above and below the plane of the ring (e.g., Chevalier 1992b). Our hydrodynamical simulations are confined to one spatial dimension and so cannot take such departures from spherical symmetry into account. However, even the most recent observations suggest that the radio source is still well inside the ring (Staveley-Smith et al. 1993) and so this should not yet be a severe restriction. Anisotropies in the radio emission itself can be attributed to relatively minor enhancements in the magnetic field strength or in the electron injection rate, which leave the overall hydrodynamic picture unaffected.

If the assumption of a spherically symmetric source is reasonable for the shock dynamics, at least until the supernova blast wave hits the ring, all that remains is to fix the location $r_\text{w}$ of the termination shock front between the freely-expanding wind and the shocked blue-giant wind. In his model of the radio emission,

Chevalier (1992a) attributes the re-emergence of radio emission in July 1990 to the arrival of the supernova blast wave at this point, which implies $r_w \approx 2 - 3 \times 10^{15}$ m, in agreement with estimates based on the strength of the blue-giant wind and the size of the cavity it has inflated inside the red-giant wind. However, this theory predicts a decrease of radio emission roughly one year after encounter, in conflict with the observed continuing increase. Thus, the precise position of the termination shock, which is is not known a priori, cannot be deduced from Chevalier's theory. Consequently, we use our model to examine two possibilities: (i) that the shock front has not yet reached $r_w$; and (ii) that the shock front passed $r_w$ some 1 – 3 years after explosion.

The calculations we present fall into two parts: the hydrodynamics of the mixture of thermal gas and cosmic rays, and the calculation of the synchrotron emission of electrons accelerated at the evolving subshock. For the first part we use the two-fluid approach (Dorfi 1990 and Duffy, Drury & Völk 1994). Since the supernova blast wave is currently in its free-expansion phase, we assume the fluid velocity behind the gas subshock to be constant in time and develop a numerical procedure for solving the equations governing the structure and evolution of the precursor. This method is described in detail in §2. Having found the hydrodynamic solution, we use it in §3 together with the model of time-dependent diffusive shock acceleration described by BK to construct the electron distribution as a function of space and time, and to find the general form of the synchrotron emission from the accelerated electrons. In §4 we compare the light curves which result from specific models with observed light curves. We conclude with a discussion of the implications of our results in §5.

## 2. The acceleration of ions by a supernova shock front

Cosmic rays (CRs) with energies up to the 'knee' of the CR spectrum (at about $10^{15}$ eV), are widely thought to be accelerated by the diffusive shock acceleration mechanism operating in supernova remnants. Recent work on this theory indicates that the supernova shock front can, whilst expanding into the interstellar medium, convert between 10% and 30% of the total blast wave kinetic energy into CRs, as required to sustain their observed intensity (Drury et al. 1989, Dorfi 1990, Jones & Kang 1990, Berezhko et al. 1993). With such high conversion efficiencies the reaction of the CRs on the gas dynamics cannot be ignored. However, most of the energy is converted into CRs during the adiabatic or 'Sedov-Taylor' phase of the evolution of the remnant, when the shock front has already overrun and set into motion a mass comparable to that ejected initially by the explosion, and is steadily decelerating as it expands further. The very young remnant of SN1987A is not in this stage of evolution. The amount of matter overrun by the shock front is very small and the ejecta are still expanding with essentially undiminished momentum. Therefore, because almost all the energy of the explosion is still tied up in the kinetic energy of the ejecta, the total amount available for acceleration of CRs is small. This does not mean the CRs are dynamically unimportant in SNR 1987A. For them to modify the shock substantially it suffices that the pressure they exert be comparable to the pressure of the post-shock gas.

We can make a simple estimate of whether or not this effect might be important by considering the rise of the CR pressure $P_C$ during the *test-particle phase*, i.e., before the CRs become dynamically important. Consider a star which explodes into a radial, constant-speed stellar wind which has an $r^{-2}$ density profile. The magnetic field of such a (highly conducting) wind has an Archimedian spiral structure (Parker 1958) which rapidly becomes azimuthal and thereafter drops off as $r^{-1}$. In the presence of a turbulent magnetic field, particles can be scattered across the shock picking up a small amount of energy each time this occurs. The timescale $t_{\rm acc}$ for acceleration is proportional to the diffusion coefficient $\kappa_p \sim \lambda_{\rm mfp} v$ where $\lambda_{\rm mfp}$ is the mean free path for scattering of a particle moving at speed $v$. Strictly speaking, $\lambda_{\rm mfp}$ refers to transport across the azimuthal field. We assume that a turbulent magnetic field of the same order as the mean field $B$ is present, and estimate the effective mean free path for transport perpendicular to the mean field (Achterberg & Ball 1994) to be the minimum indicated by the quasilinear theory, i.e., the particle gyroradius $r_g \propto p/B$. Therefore, the relatively strong magnetic field near the surface of the progenitor of SN1987A (Kirk & Wassmann 1992) implies very rapid acceleration to high energies (Völk & Biermann 1988). For this rough estimate of $P_C$ we assume the cosmic rays are ultra-relativistic, neglect the effects of adiabatic expansion on their energy and neglect also the time dependence of the magnetic field at the shock front. It is possible to relax these assumptions, but doing so produces no qualitative change in the general picture. Assuming that acceleration starts at time $t_0$, we calculate the upper energy cutoff from $\dot{p}_{\rm max} = p_{\rm max}/t_{\rm acc} \propto B/v$. The momentum distribution of the CRs is then just that for diffusive acceleration at a strong shock, $N(p) \propto p^{-2}$. It follows that the temporal dependence of the CR pressure is given by

$$P_C \propto \frac{Q(t)}{A(t)} \int_{p_0}^{p_{\rm max}(t)} \frac{pc}{3} \left(\frac{p}{p_0}\right)^{-2} dp \qquad (1)$$

where $Q(t)$ is the rate at which particles are injected with momentum $p_0$ and $A(t)$ is the area of the shock front. For a freely-expanding spherical shock, $A \sim t^2$ and if we assume a constant fraction $\eta$ of the overtaken thermal particles is injected, then $Q \propto \eta n A$ is constant for an $n \propto r^{-2}$ density gradient. The CR pressure then varies according to $P_C \sim t^{-2} \ln(t/t_0)$. Whereas the post shock gas pressure, $P_G$, falls of as $t^{-2}$, $P_C$ will initially ($t < 1.7 t_0$) increase very rapidly on a timescale much shorter than the expansion timescale of the shock, $r_s/\dot{r}_s$, where $r_s(t)$ is the radius of the shock. When $t > 1.7 t_0$ the CR pressure decreases with time, but it only decreases as rapidly as the gas pressure in the asymptotic limit, $t \gg t_0$. It is therefore plausible that $P_C$ becomes the dominant component of the total pressure during the early phase of the remnant's evolution, depending on the values of the relevant parameters such as $\eta$ and $\kappa_p$. Once this happens the test particle estimates break down and it becomes necessary to solve the self-consistent evolution numerically.

## 2.1. The Two-Fluid Treatment

We develop a numerical solution by considering a two-fluid system consisting of thermal gas and CRs. All the mass is assumed to reside in the gas, but the total pressure is the sum of $P_G$ and $P_C$. In this picture there are three processes that can change $P_C$ in a frame comoving with the fluid. The first is compression: the CRs are tied to the fluid through the magnetic field and, therefore, change their energy when the fluid flow expands or contracts. To calculate this effect we need the CR equation of state, $P_C = (\gamma_C - 1) E_C$, where $E_C$ is the CR energy density and $\gamma_C$ the adiabatic index which decreases from 5/3 to 4/3 as ultrarelativistic particles begin to dominate the spectrum. However, without knowledge of the spectrum we cannot calculate $\gamma_C$ exactly. A similar problem applies to the second process that can modify $P_C$: the diffusion of CRs off magnetic irregularities, because, in general, the diffusion coefficient depends on the individual particle energy. Working on a macroscopic level with fluid quantities, we replace $\kappa_p(p)$ by an average quantity $\overline{\kappa}_p$ weighted over the (unknown) energetic particle spectrum. The simplification to a two-fluid model is thus achieved at the price of introducing two undetermined closure parameters: $\gamma_C$ and $\overline{\kappa}_p$. In the present case the initial rapid acceleration will quickly lower $\gamma_C$ to 4/3, and we use this lower limit for the CR adiabatic index. Our calculations therefore underestimate $P_C$ relative to $E_C$, providing a lower limit on the amount of shock modification. For $\overline{\kappa}_p$ we adopt the model used in other two-fluid treatments of CR acceleration (see Drury, Markiewicz and Völk 1989 and Duffy, Drury and Völk 1994), namely $\overline{\kappa}_p(r) = \kappa_p(p_{max})/4$ and $\kappa_p(p_{max}) = p_{max} c/3qB(r)$. This prescription results in the required CR conversion efficiencies at older SNRs, and the strong X-ray luminosity of young SNRs (Dorfi 1990).

The third process which affects the CR pressure is that of injection. Because the two-fluid model does not deal with the number density of CR particles, but only with their energy density, injection is modelled by transferring a small fraction $\epsilon$ of the available mechanical energy at the subshock into CRs. The energy flux of injected particles is then given by $F_i = 0.5\epsilon \left[\rho_1(U_s - U_1)^3 - \rho_2(U_s - U_2)^3\right]$, where $\rho_{1,2}$ and $U_{1,2}$ are the density and velocity, respectively, immediately upstream and downstream of the shock. This prescription naturally implies a lower injection efficiency for weak shocks. The momentum of freshly injected CRs, which is needed for the computation of the maximum momentum, is taken to be a multiple, $\lambda$, of the post-shock gas sound speed $c_s$ times the proton mass $m$, i.e., $p_0 = \lambda m c_s$. In the following, we use values for $\epsilon$ and $\lambda$ consistent with those needed to explain the observed galactic CRs, assuming an SNR origin, namely, $\lambda = 2$ and $\epsilon = 1 - 5 \times 10^{-3}$.

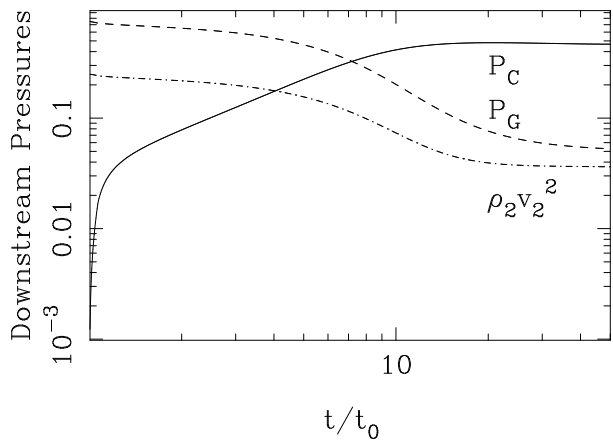

Fig. 1.— The time-dependence of the cosmic ray pressure $P_C$ (solid line), the gas pressure $P_G$ (dashed line) and the momentum flux $\rho_2 v_2^2$ of matter leaving the shock (dot-dashed line) with an injection level of $\epsilon = 5 \times 10^{-3}$. All are normalised to the upstream ram pressure of the freely-expanding wind, $\rho_G U_s^2$, which falls off as $r_s^{-2}$.

In the free-expansion phase of an SNR, there are two shock fronts present: an outer or forward shock entering the progenitor's wind, and a reverse shock moving into the ejecta. The magnetic field immediately behind the forward shock is just the compressed field of the wind, which we expect to be predominantly toroidal and inversely proportional to the shock radius $r_s$, at least out to the beginning of the stagnation zone. At constant shock speed, therefore, the magnetic field immediately behind the outer shock falls off as $t^{-1}$. On the other hand, the magnetic field in the homologously expanding ejecta varies as $B \propto r^{-2}$. This is the field encountered by the reverse shock, which moves outwards at a speed slightly lower than that of the ejecta. Consequently, the field immediately behind the reverse shock decays as $t^{-2}$, much more rapidly than that behind the forward shock. We are primarily interested in the synchrotron radiation from SNR 1987A, and so confine our attention to particles accelerated in the stronger field at at the outer shock; all other things being equal, the synchrotron emission from the reverse shock should not be as bright as that from the forward shock.

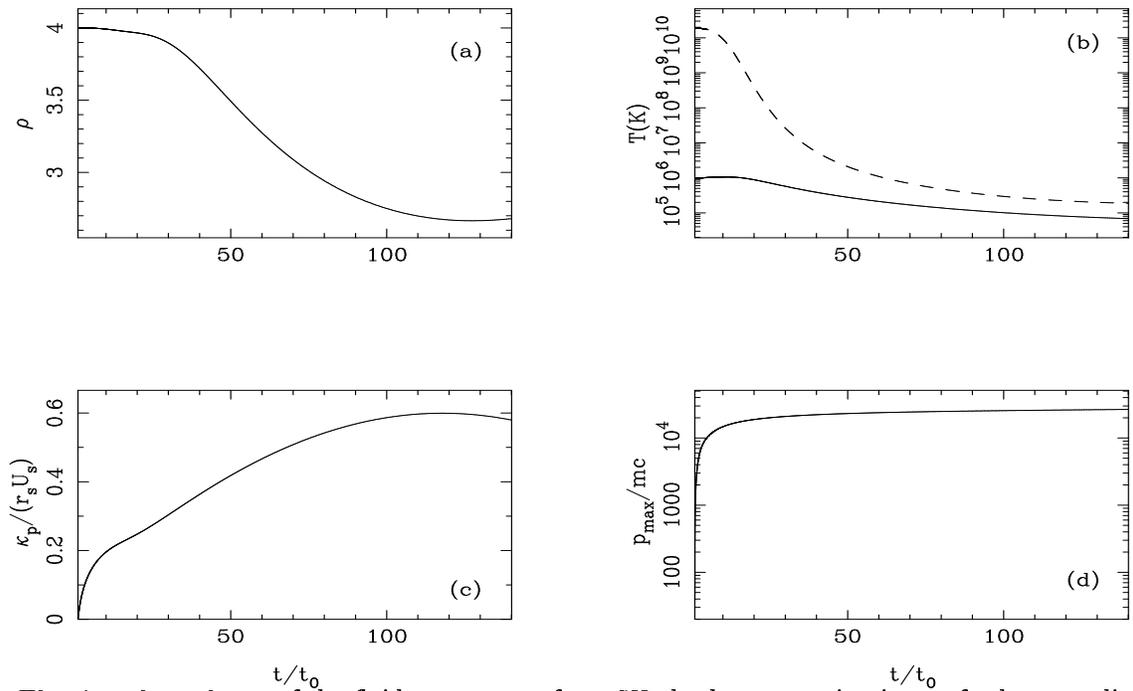

Fig. 2.— The time dependence of the fluid parameters for a SN shock propagating into a freely-expanding wind, for the same parameters used to produce Figure 1. The panels show: (a) the subshock compression ratio, (b) the pre- (solid line) and post (dashed line) subshock temperatures, (c) the weighted proton diffusion coefficient $\bar{\kappa}_p$ in units of $r_s U_s$, and (d) the maximum CR momentum in units of $mc$.

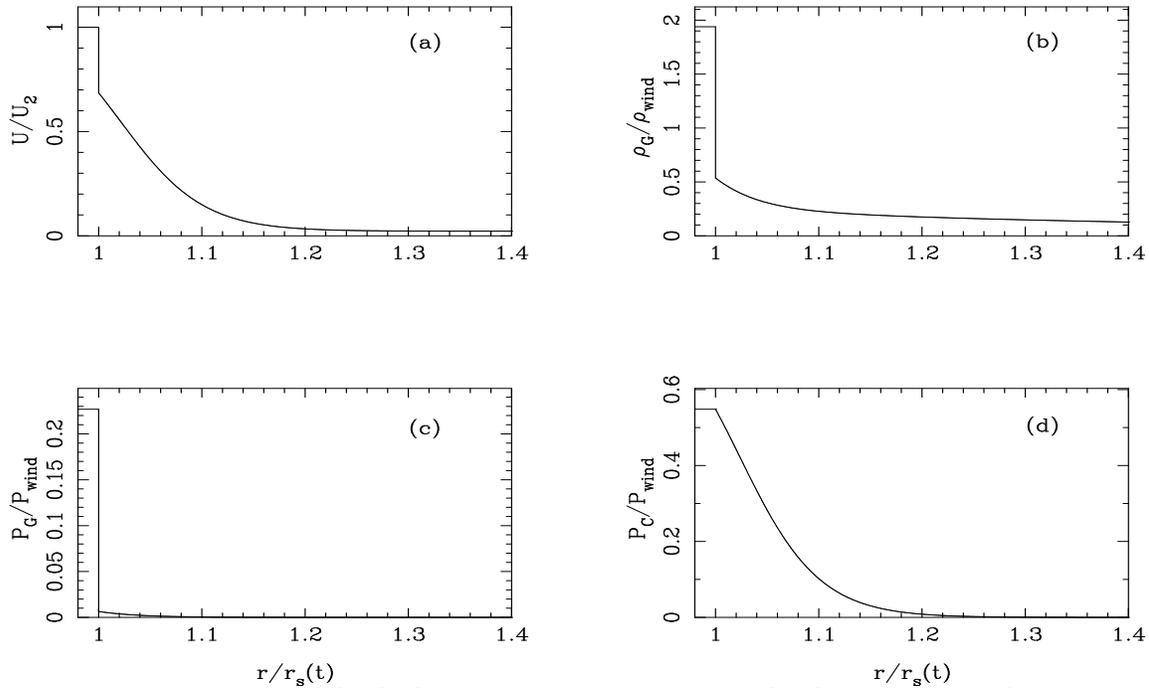

Fig. 3.— The spatial dependence of the fluid parameters at time $t = 30 t_0$ for the case presented in Figures 1 & 2. The panels show: (a) the fluid velocity $U$, (b) the gas density $\rho_G$, (c) the gas pressure $P_G$, and (d) the CR pressure in the precursor $P_C$.

To account for the dynamics of the ejecta in the free-expansion phase, we assume the flow velocity of the gas immediately behind the outer shock is constant in time. In fact, in spherical geometry, the speed of the shocked gas is not exactly equal to the constant speed of the ejecta, which constitute a piston driving the flow. This is because the pressure behind the outer shock drops off roughly as $r_s^{-2}$. If the shocked gas were to maintain constant speed, adiabatic expansion with $n \propto r^{-2}$ would cause a more rapid decline ($P_G \propto r^{-8/3}$ for expansion of a gas of adiabatic index 5/3). Instead, the shocked gas settles slowly onto the piston. We neglect the small velocity difference this implies and solve the coupled equations of CR hydrodynamics upstream of the shock. In this particular problem, it is important to have good spatial resolution of the precursor. We therefore use a formulation in which the spatial grid expands along with the shock front. Hereafter we use upper case $U$'s to denote fluid speeds in the rest frame of the explosion's centre, and lower case $v$'s for fluid speeds in the rest frame of the subshock. Defining $\xi = r/r_s(t)$, the coupled two-fluid equations in the upstream region can be written in the form

$$\frac{D\rho_G}{Dt} = -\frac{\rho_G}{r_s \xi^2} \frac{\partial}{\partial \xi}\left(\xi^2 U\right) \quad (2)$$

$$\frac{DU}{Dt} = -\frac{1}{r_s \rho_G} \frac{\partial}{\partial \xi}(P_G + P_C) \quad (3)$$

$$\frac{DS}{Dt} = 0 \quad (4)$$

$$\frac{DP_C}{Dt} = -\frac{\gamma_C P_C}{r_s \xi^2} \frac{\partial}{\partial \xi}\left(\xi^2 U\right)$$
$$+ \frac{1}{(r_s \xi)^2} \frac{\partial}{\partial \xi}\left(\xi^2 \overline{\kappa}_p \frac{\partial P_C}{\partial \xi}\right)$$
$$+ \frac{F_i}{r_s}(\gamma_C - 1)\,\delta(\xi - 1) \quad (5)$$

where $U$ is the bulk plasma flow speed in the rest frame of the explosion, and where

$$\frac{D}{Dt} \equiv \frac{\partial}{\partial t} + \left(\frac{U - \xi U_s}{r_s}\right) \frac{\partial}{\partial \xi} \quad (6)$$

with $U_s$ the speed of the gas subshock. The first two of these equations describe the conservation of mass and momentum of the gas, with the CR pressure gradient included as an additional force; $\rho_G$ is the gas density. Equation (4) describes the adiabatic evolution of the gas in the precursor region; $S$ is the entropy per unit mass. We have assumed there is no dissipation of energy into the gas, i.e., that there is no damping of the Alfvén waves responsible for the scattering of the cosmic rays. Although it is possible to take account of this effect, we expect it to be unimportant when, as in this case, the direction of the average field is normal to the cosmic ray pressure gradient (Jones 1992). Equation (5) describes the processes which affect the CR fluid. The three terms correspond to compression, diffusion and injection, respectively. Finally, in order to compute the effective diffusion coefficient $\overline{\kappa}_p$, we need the magnetic field intensity as a function of position and time. Before explosion, the constant velocity stellar wind contained a toroidal field $B \propto 1/r$, up to the beginning of the stagnation zone at $r_w$, after which $B \propto r$. On moving into this configuration, the shock front forms a precursor, which sets the plasma ahead of the shock into motion, compressing both it and its frozen-in magnetic field. Assuming this remains toroidal, we find that $B$ is governed by the equation

$$\frac{DB}{Dt} = -\frac{B}{r_s \xi} \frac{\partial}{\partial \xi}(\xi U) \quad . \quad (7)$$

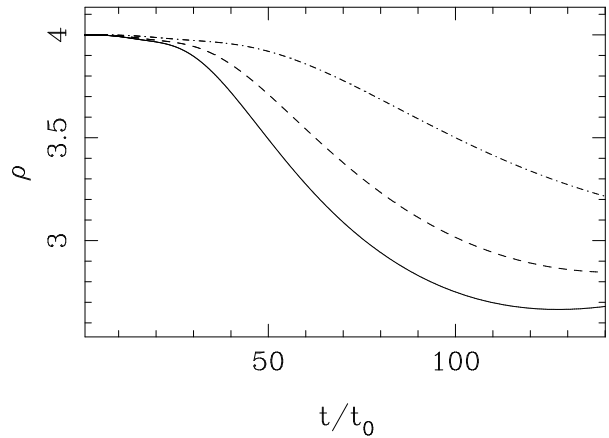

Fig. 4.— Evolution of the subshock compression ratio for a SN shock propagating into a freely-expanding wind, with CR injection levels of $\epsilon = 0.005$ (solid line), 0.003 (dashed line) and 0.001 (dot-dashed line).

The CR pressure gradient at the shock is obtained by integrating (5) across the discontinuity at $\xi = 1$ to give

$$\left.\frac{\overline{\kappa}_p}{r_s}\frac{\partial P_C}{\partial \xi} - \gamma_C P_C U\right|_{1-\epsilon}^{1+\epsilon} = -F_i(\gamma_C - 1) \quad (8)$$

which is then used as a boundary condition. All that remains is to determine the speed of the gas subshock.

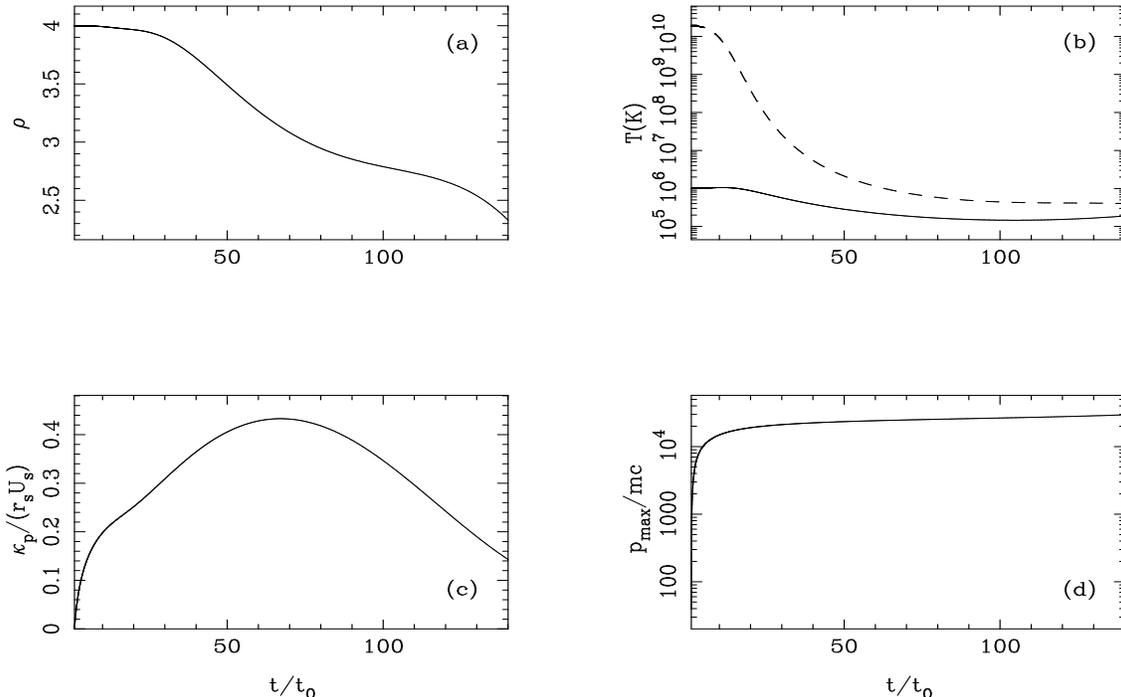

Fig. 5.— The time dependence of the fluid parameters for a SN shock propagating into a confined wind. The termination shock is at a radius $r_{\rm w} = 67 U_2 t_0$. Otherwise, the wind parameters are as used to produce Figure 1. The panels show: (a) the subshock compression ratio, (b) the pre- (solid line) and post (dashed line) subshock temperatures, (c) the weighted proton diffusion coefficient $\overline{\kappa}_{\rm p}$ in units of $r_s U_s$, and (d) the maximum CR momentum in units of $mc$.

This can be obtained from the Rankine-Hugoniot relations, extended to include the extracted energy flux $F_{\rm i}$. The upstream flow velocity in the rest frame of the subshock $v_1$ is then given by the quadratic equation

$$v_1^2 - v_1 \frac{\gamma_{\rm eff}+1}{2}(U_1 - U_2) - \frac{\gamma_{\rm eff} P_1}{\rho_1} = 0 \qquad (9)$$

with $\gamma_{\rm eff} = \gamma_{\rm G}/[1 + \epsilon(\gamma_{\rm G} - 1)]$, where $P_1$ is the gas pressure just upstream of the subshock. ¿From $v_1$, we then obtain the subshock velocity, $U_{\rm s} = U_1 - v_1$. The details of the numerical scheme used to solve these time-dependent equations are given in Appendix A.

## 2.2. The Freely-Expanding Wind

We solve the two-fluid equations with initial conditions believed to be relevant to the environment around the star Sanduleak $-69°202$, the progenitor of SN1987A (Chevalier & Fransson 1987, Storey & Manchester 1987). Consider first the freely-expanding blue supergiant wind moving with a constant speed of $500\,\rm km\,s^{-1}$. Together with a mass-loss rate of $10^{-5}\,\rm M_\odot\,yr^{-1}$ and a stellar radius of $r_* = 3 \times 10^{10}$ m, this determines the proportionality constant of the $r^{-2}$ density profile. The undisturbed magnetic field is set to $B = r_* B_*/r$ where $B_* = 2\,\rm mT$ (20 G) (see Kirk & Wassmann 1992). When the star exploded it ejected about $1\,\rm M_\odot$ at a speed of some $22,500\,\rm km\,s^{-1}$; this then is the speed of the downstream medium $U_2$ that is kept constant in our solutions for the free expansion of the remnant. The two parameters controlling injection: $\epsilon$ and $\lambda$ (the latter appears only in the calculation of the effective diffusion coefficient $\overline{\kappa}_{\rm p}$) are set to $5 \times 10^{-3}$ and 2 respectively.

The only remaining unknown is the time $t_0$ at which CR injection begins. At very early times, we expect the strong magnetic field to be very close to its unperturbed spiral form. Further out, small fluctuations in the speed of the wind of the progenitor are likely to have produced a more turbulent field. In addition the energetic particles themselves can generate turbulence. We assume that the scattering of CRs and injection becomes efficient when the magnitude of the fluctuations in the magnetic field approaches the average field magnitude, but the time at which this is likely to occur is essentially unknown. We therefore take $t_0$ as a free parameter. However, as we show in Appendix A, $t_0$ introduces the only physical timescale into the problem, and the time-dependent evolution we compute scales precisely with this time.

The evolution of $P_{\rm C}$, $P_{\rm G}$ and $\rho_2 v^2$ downstream of the subshock (normalised to the upstream ram pressure of the freely-expanding wind, $\rho_{\rm G} U_{\rm s}^2$ which falls off as $r_{\rm s}^{-2}$), are shown in Figure 1. $P_{\rm C}$ exhibits a rapid initial rise, followed by a decline in absolute terms which is somewhat slower than the $\sim t^{-2}$ fall off of $P_G$ and of $\rho_G U_s^2$. This behaviour is in accordance with our earlier test-particle estimate, and is characteristic of shock acceleration of CRs in a stellar wind cavity. For the parameter values used here the CR pressure eventually dominates the thermal pressure at $t \gtrsim 10 t_0$ and the gas subshock starts to weaken.

This effect is depicted in Figure 2, which shows the evolution of the remaining fluid parameters. Panel (a) shows that after about $100 t_0$ the subshock compression ratio drops to a level which is comparable to that suggested by BK's theory for the radio emission of SNR 1987A (i.e., electron acceleration at the gas subshock). The remaining panels show the temperatures of the pre- and post-shock fluids, the weighted diffusion coefficient $\overline{\kappa}_p$ in units of $r_s U_s$ and the evolution of the upper CR momentum cut-off $p_{max}$ with adiabatic losses included.

Figure 3 shows the spatial dependence of the hydrodynamic solution at time $t = 30 t_0$. It is noteworthy that the precursor reaches a size which is comparable to the radius of the subshock. A simple estimate based on the equation of motion (3) shows that the subshock compression ratio can be significantly modified (i.e., $P_C$ can be of the order of $\rho_G U_s^2$) only if the size of the precursor is a significant fraction of the radius of the shock. As $P_C$ becomes dynamically important the compression ratio between points upstream of the precursor and downstream of the subshock increases from 4 to values typically greater than 10 (fig. 3b) since the diffuse, relativistic protons alter the overall equation of state.

The time dependence of the subshock compression ratio is sensitive to the injection parameter $\epsilon$. This is shown in Figure 4. The higher the level of CR injection, the more rapidly the subshock weakens, and the lower the value of $\rho$ at the plateau.

### 2.3. The Confined Wind

The second situation we investigate consists of a SN shock expanding into a bubble of freely-expanding wind which is terminated at a shock front, outside which there is a constant density stagnation zone. In the stagnation zone the velocity of the circumstellar material decreases as $r^{-2}$, and the toroidal magnetic field increases as $r$. In order to simulate this situation, we set up an initial density profile which contains a sudden transition from $n \propto r^{-2}$ to $n = $ constant at a radius $r_w$. The other parameters are as used for the freely-expanding wind case discussed above. Once again, the time at which injection starts is the only physical timescale, provided $r_w$ is expressed as a multiple of the velocity of the downstream plasma $U_2$ times $t_0$.

The time dependence of the fluid parameters in this case is shown in Figure 5. We have chosen to place the termination shock at a radius $r_w = 67 U_2 t_0$. The most important difference between the solution shown in Figure 5 and the freely-expanding wind solution shown in Figure 2 is the decrease of the effective diffusion coefficient $\overline{\kappa}_p$ brought about by the increase in $B$ in the stagnation zone. This leads to a characteristically different evolution of the subshock compression ratio, in which there are two points of inflection separated by a more or less pronounced plateau region. In the freely-expanding wind case (Figure 2), only one point of inflection is seen.

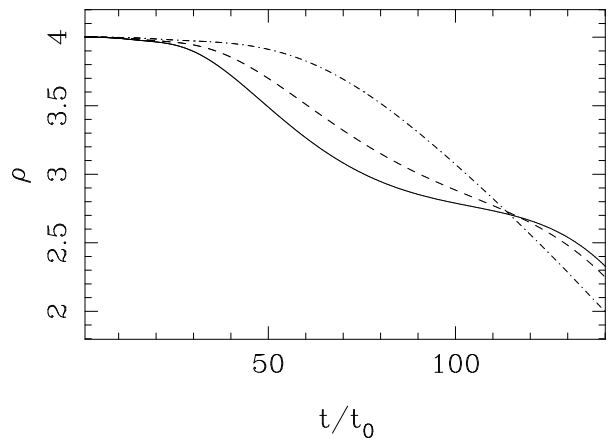

Fig. 6.— The effect of the CR injection level on the evolution of the subshock compression ratio, for the confined wind case. The results for $\epsilon = 0.005$ (solid line), 0.003 (dashed line) and 0.001 (dot-dashed line) are shown.

This can be seen in more detail in Figure 6 which shows the effect of the injection parameter $\epsilon$ on the evolution of the subshock compression ratio. At high injection rates, the plateau at a compression ratio of about 2.7 is quite persistent.

The position of the termination shock also has a strong effect on the evolution of the subshock compression ratio, as can be seen in Figure 7. The solid line in Figure 7 shows that after a period where the subshock compression ratio is essentially constant – the 'plateau' discussed earlier – the subshock begins to weaken again and eventually disappears. This is a general feature of all the solutions for which there is substantial CR modification of the shock. The dot-dashed line in Figure 7 shows that placing the stagnation zone very close to the star results in the rapid smoothing out of the subshock without any significant plateau phase.

### 3. Acceleration of electrons

The pressure of the ultrarelativistic electrons responsible for the observed synchrotron emission from SNR 1987A is of the order of $10^{-8}$ Pa ($6 \times 10^4$ eV cm$^{-3}$, Ball & Kirk, 1992b), which is only a few percent of the cosmic ray pressure. Electrons which are accelerated by the shock therefore make no contribution to the hydrodynamics, and can be treated as test particles. Furthermore, the electron diffusion lengthscale, at GeV energies, inferred from earlier model fits to the observations, $\kappa_e/U_s \sim 10^{13}$ m (BK), is much smaller than the precursor lengthscale which is determined by $10^4$ GeV protons, of $\sim 10^{15}$ m obtained in section 2 (see Figure 3). Therefore electrons in the upstream fluid which are overtaken by the shock will initially experience the shock precursor as a slow compression, then see the gas subshock as a simple, sudden discontinuity in the fluid speed, and subsequently pass into the slowly expanding downstream fluid. Diffusive shock acceleration of electrons will thus occur at the gas subshock, which has a much lower compression ratio than that of the subshock-precursor system responsible for accelerating the CRs.

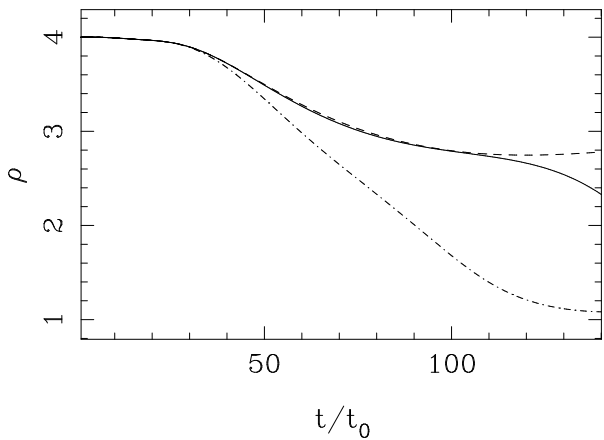

Fig. 7.— The effect of the the position of the termination shock $r_w$ on the evolution of subshock compression ratio. The values of $r_w$ are 100, 67 and 45 times $U_2 t_0$ (dashed, solid and dot-dashed lines respectively).

Our treatment of the electron acceleration involves a straightforward generalisation of that by BK, including now the effect of the evolution of the shock strength with time. The two important physical processes which modify the electron distribution – acceleration and adiabatic expansion – are treated separately. We assume that the acceleration occurs close to the shock on a timescale which is short compared to the expansion timescale $r/U_s$. We therefore first model the effect of diffusive acceleration in the vicinity of the shock without adiabatic losses, and then, once the accelerated particles escape downstream of the shock, the flow is assumed to be diffusion-free and adiabatic expansion losses are included.

In the rest frame of the shock a spatially-averaged model of diffusive acceleration leads to the equation

$$\frac{\partial N}{\partial t} + \frac{\partial}{\partial p}\left(\frac{p\Delta}{t_c}N\right) + \frac{P_{\rm esc}}{t_c}N = Q(t)\,\delta(p-p_0) \quad (10)$$

where, for a particle of speed $v$, $t_c = 4\kappa(1/v_1 + 1/v_2)/v$ is the average time taken to cross and recross the shock, $\Delta = 4(v_1 - v_2)/3v$ is the average fractional increase of particle momentum per cycle, and $P_{\rm esc} = 4v_2/v$ is the probability per cycle of a particle being advected downstream away from the shock. Plasma flows into the shock at speed $v_1$ and out of it at speed $v_2$. Equation (10) can be solved by the method of characteristics. We assume, for simplicity, that the electron diffusion coefficient $\kappa_e$ (and hence $t_c$) is independent of momentum. This is valid for the modelling of the radio emission, because the narrowness of the "radio window" means that the electrons which contribute to the radio emission span less than a decade in momentum, over which the variation in $\kappa_e$ is relatively small. We also assume that the electron injection rate, $Q(t)$, is zero up until some time $t_a$. The solution to equation (10) can then be written in integral form, the details of which are given in Appendix B. When the shock properties are independent of time and the injection rate is constant after being switched on at $t_a$, the general solution reduces to that of BK, namely the general solution reduces to that of BK, namely

$$N(p,t) = \frac{t_c Q}{p_0 \Delta}\left(\frac{p}{p_0}\right)^{-2\alpha-1} \times$$
$$\left[H(p-p_0) - H\left(p - p_0 e^{(t-t_a)\Delta/t_c}\right)\right] \quad (11)$$

where $\alpha = P_{\rm esc}/(2\Delta) = 3/[2(\rho - 1)]$ and $\rho$ is the (constant) shock compression ratio. On the other hand, when the acceleration takes place at a subshock which is evolving due to the CR modifications, $\Delta$, $t_c$ and $P_{\rm esc}$ are all functions of time. The general solution to equation (10) can then be integrated numerically along the characteristics determined by the hydrodynamic solution for the shock evolution. The qualitative features of a typical accelerated electron distribution resulting from a CR-modified shock, which is weakening with time, will then be as follows. At momenta just above that of injection, $p_0$, the spectrum will be a power law as determined by acceleration at a simple shock of the appropriate instantaneous compression ratio $\rho(t)$. That is, for some range of $p$ just above $p_0$, $N(p,t) \propto p^{-q}$ where $q = (\rho(t)+2)/(\rho(t)-1)$. At higher momenta the spectrum will be harder than that from a shock of constant compression ratio $\rho(t)$, reflecting the fact that the modified shock was stronger at earlier times. Spectra with just such a concave shape are found when equation (10) is integrated numerically using the hydrodynamic solutions of section 2. Figure 8 shows the spectrum which results after 5 years in such a case where electron injection is switched on after 2.5 years and then remains constant, with 1% of electrons swept up by the subshock injected into the acceleration process. The value of the electron diffusion coefficient is taken to be $\kappa_e = 10^{16}\,\mathrm{m^2\,s^{-1}}$. The evolution of the hydrodynamics is that of the freely-expanding wind solution of Figure 2 with $t_0 = 18.3$ days. The solid line in Figure 8 is the electron spectrum multiplied by $p^{q(5\,\mathrm{yr})}$ with $q(5\,\mathrm{yr}) = 2.52$ corresponding to the instantaneous subshock compression ratio of $\rho(5\,\mathrm{yr}) = 2.98$. The fact that the solid line in Figure 8 is essentially horizontal near $p_0$ indicates that the spectrum of the lowest energy electrons is determined by the instantaneous subshock compression ratio. The electron spectrum at the upper momentum cut-off is harder, with $q = 2.22$, and is related to the subshock compression ratio at the time when electron injection began $\rho(t_a) = 3.46$ (dashed line). This follows from the fact that the electrons with momenta near the maximum are those which have spent the longest possible time undergoing acceleration, and so must have been injected when electron injection began.

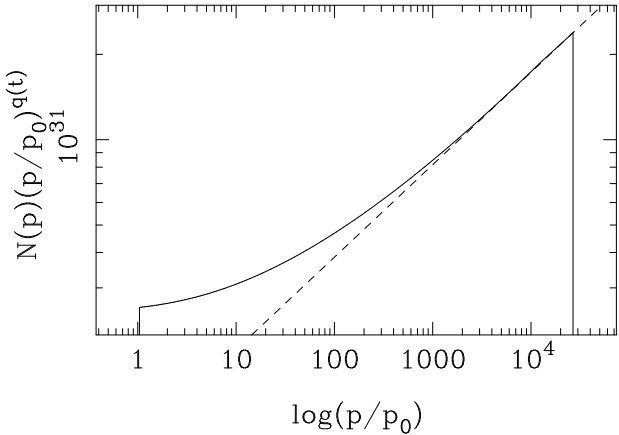

Fig. 8.— Curvature of the electron spectrum produced by acceleration at the CR-modified subshock of Figure 2, 5 years after the explosion, assuming electron injection began at $t_a = 2.5$ years. The spectrum has been multiplied by $p^{q(5\,\mathrm{yr})}$ (solid line) to show that the most recently injected particles, i.e., those near $p_0$, have a spectrum determined by the instantaneous subshock compression ratio. The slope of the spectrum near the upper cut-off is compared with the spectral index given by the subshock compression ratio at $t_a$ (dashed line).

The second process affecting the electrons is escape downstream followed by adiabatic losses in the expanding fluid. Electrons escape from the shock at a rate of $N(p,t)P_{\mathrm{esc}}/t_c$ per second, and plasma leaves the shock at a speed $v_2$. If the shock is accelerating electrons over an area $A$ it follows that the distribution function of the escaping electrons is

$$f_s(p,t) = \frac{1}{4\pi p^2} N(p,t) \frac{P_{\mathrm{esc}}}{A t_c v_2} \qquad (12)$$

where the subscript s indicates that this applies only immediately downstream of the shock. The divergence of the downstream flow modifies the distribution function which is therefore of the form $f(p,\mathbf{r},t)$ and satisfies

$$\frac{\partial f}{\partial t} + \mathbf{v}\cdot\nabla f - \frac{1}{3}\nabla\cdot\mathbf{v}\,p\frac{\partial f}{\partial p} = 0 \qquad (13)$$

where $\mathbf{v}$ is the flow velocity of the downstream fluid. For radial flow, $\mathbf{v} = U_2 \hat{\mathbf{r}}$, this equation may be solved using the method of characteristics after introducing the Lagrangian (comoving) coordinate $\mathbf{R}(\mathbf{r},t)$ which is defined implicitly by the equation

$$\mathbf{r} = \mathbf{R}(\mathbf{r},t) + \int_0^t dt'\,\mathbf{v}(\mathbf{R},t') \qquad (14)$$

so that $\mathbf{R} = \mathbf{r}$ when $t = 0$. This coordinate is constant for a given mass element in the downstream fluid. The details of the solution are again given in Appendix B, and the general result is

$$f(p,R,t) = \frac{1}{4\pi(xp)^2}\frac{P_{\mathrm{esc}}}{A t_c v_2} \times$$
$$N\left(p[(R+U_2 t)/(\rho R)]^{2/3}, t\right) \qquad (15)$$

where $x = [(R+U_2 t)/(\rho R)]^{2/3}$ and $v_2 = U_s - U_2$.

Since SNR 1987A is still in its free-expansion or piston-driven phase, the material behind the shock is being pushed out by the ejecta at a speed which is essentially constant. There is a slight variation in the downstream speed with radius, decreasing marginally from the contact discontinuity to the position of the shock, but this can reasonably be neglected here, and has been neglected in the two-fluid model of section 2. The relationship between the downstream electron distribution $f(p,R,t)$ and the distribution of accelerated electrons at the shock front $N(p,t)$ is therefore unchanged by the inclusion of the cosmic ray modifications of the shock front. The only effect of the modifications is that on $N(p,t)$ itself.

Integrating over the downstream electron distribution gives their synchrotron flux density at a distance $D$ from the source synchrotron flux density at a distance $D$ from the source

$$F_d(\nu,t) = \frac{4\pi}{D^2}\int_0^\infty dp\,p^2 \int_{R_a(t)}^{R_s(t)} dR$$
$$(R+U_2 t)^2 j_\nu(p) f(p,R,t) \qquad (16)$$

where $R_a(t) = R[r_s(t_a), t_a]$ is the Lagrangian coordinate of the first injected particles, $R_s(t) = R[r_s(t), t]$

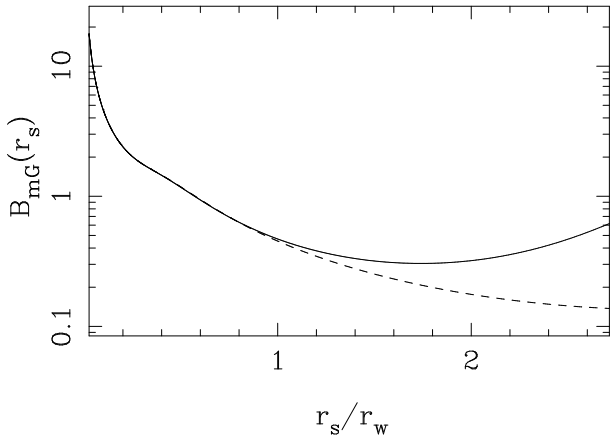 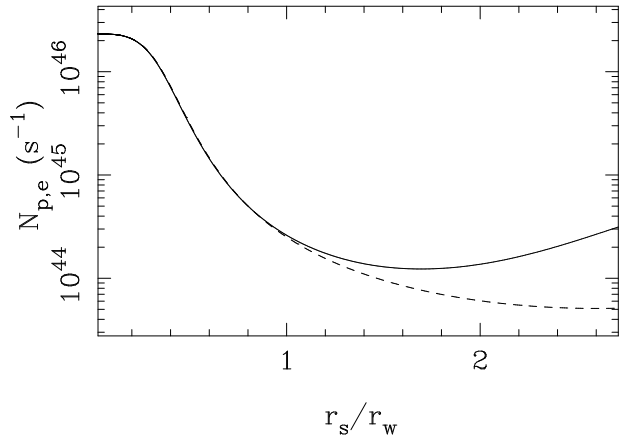

Fig. 9.— The evolution of the pre-subshock magnetic field (in mG) at a CR-modified shock propagating into a freely-expanding stellar wind (dashed line), and into a confined wind which has a termination shock at radius $r_w$ surrounded by a stagnation zone (solid line).

Fig. 10.— Evolution of the rate at which protons (and electrons) enter the subshock, for the freely-expanding stellar wind case (dashed line), and the confined wind case with a termination shock at radius $r_w$ surrounded by a stagnation zone (solid line).

is the Lagrangian coordinate of the most recently injected particles, and $j_\nu(p)$ is the single-particle synchrotron emissivity. There is also a contribution to the emitted flux from the accelerating electrons which are still in the vicinity of the shock, which is given by

$$F_s(\nu, t) = \frac{1}{4\pi D^2} \int_0^\infty dp\, j_\nu(p) N(p, t). \quad (17)$$

To simplify the calculation of the flux density we use $j_\nu(p) = a_0 (p/mc)^2 B^2 \delta[\nu - a_1(p/mc)^2 B]$ with $a_0 = 1.6 \times 10^{-14}$ W Hz$^{-1}$ T$^{-2}$ and $a_1 = 1.3 \times 10^{10}$ Hz T$^{-1}$. The evolution of the flux density with time will depend on the details of the model but some general comments are appropriate here. Emission at a particular frequency $\nu$ will switch on when there are electrons with Lorentz factor $\gamma_L$ which satisfy the condition $\gamma_L = [\nu/a_1 B(r,t)]^{1/2}$. Therefore, emission will appear earlier at low frequencies than at high frequencies because of the finite acceleration time: emission at higher frequencies requires the presence of electrons with higher energies (or a higher $B$). After switch, on the flux density at a fixed frequency will initially increase with time as more electrons are accelerated up to the required energy. On a somewhat longer timescale adiabatic energy losses by the electrons become important. If the magnetic field (or the rate of injection) decreases with radius then emission from the newly accelerated electrons will not be able to compensate for the adiabatic energy losses of the electrons which have already escaped from the shock into the downstream plasma. As a result the flux density at a given frequency will peak, and then begin to fall.

The evolution of the emission due to acceleration at a SN shock propagating into, on the one hand, an undisturbed freely-expanding stellar wind, and, on the other, a confined wind, is quite different. We now examine these two situations using the hydrodynamic solutions presented in Figure 2 and Figure 5 respectively.

The evolution of the magnetic field immediately upstream of the subshock, in these two cases, is shown in Figure 9. As the shock propagates through the undisturbed wind the pre-subshock magnetic field initially falls off as $r_s^{-1}$. This is apparent in Figure 9 at shock radii $r_s \lesssim 0.2 r_w$. When the CR effects become dynamically important, the CRs in the precursor begin to accelerate the stellar wind material ahead of the subshock. This compresses the magnetic field, and leads to a decline in the pre-subshock field which is somewhat slower than $r_s^{-1}$. In the confined wind, the undisturbed field dependence changes at $r_w$ from $B \propto r^{-1}$ to $B \propto r$, but since the magnetic field is assumed to be frozen into the plasma, the evolution of the field at the subshock is very much smoothed out by the modification of the upstream plasma by the CR precursor. Thus the confined wind case gradually evolves away from the freely-expanding wind case, starting when the precursor encounters the termination shock (at $r_s \approx 0.9 r_w$ in Figure 9) and eventually, when $r_s$ is considerably larger than $r_w$, the pre-subshock field increases with increasing $r_s$ (time).

The rate at which protons (and electrons) are swept up by the subshock is important because we assume a certain fraction of these is injected into the diffusive acceleration process. The rate goes through the same three phases – an unmodified wind phase, a wind phase in which CR effects are important, and (if applicable) a stagnation zone phase. The evolution of the total number of protons (and electrons) crossing the subshock per unit time, $N_{p,e}$, is shown in Figure 10 – again for the two hydrodynamic solutions presented in Figures 2 and 5. Initially, the shock expands at constant speed, its area increases as $r_s^2$, and since the density of the undisturbed wind decreases as $r^{-2}$ the rate at which plasma crosses the subshock is constant ($r_s \lesssim 0.2 r_w$ in Figure 10). Then, as CR effects become important the plasma upstream of the subshock is accelerated, and $N_{p,e}$ decreases as the shock expands further. In the confined wind case there is a gradual turnup in $N_{p,e}$, starting when the precursor encounters the termination shock ($r_s \approx 0.9 r_w$ in Figure 10).

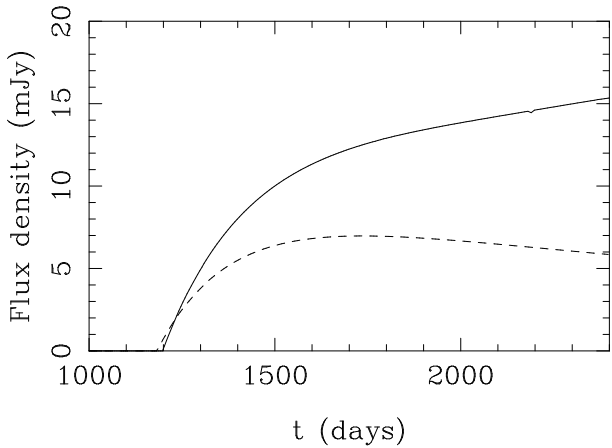
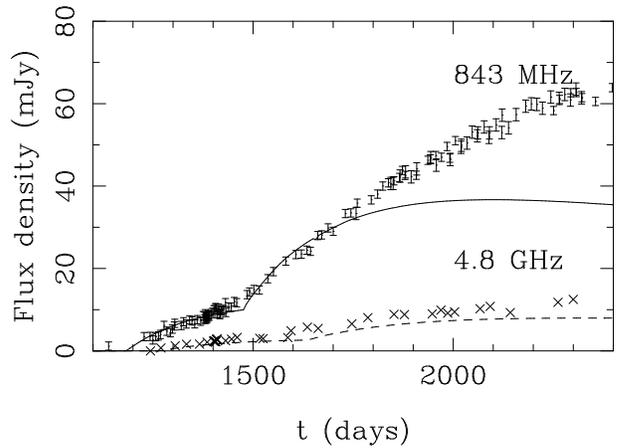

Fig. 11.— Evolution of the flux density from electrons accelerated by a CR-modified SN shock propagating into a freely-expanding stellar wind (dashed line), and into a confined wind which has a termination shock at radius $r_w = 2 \times 10^{15}$ m surrounded by a stagnation zone (solid line). In both cases 1% of electrons incident upon the shock are injected into the acceleration process.

Fig. 12.— 'Best' model fit to the observed radio emission from SNR 1987A, for acceleration at a CR-modified shock propagating into a freely-expanding stellar wind. For clarity only a subset of the data at 843 MHz is shown. The uncertainties in the 4.8 GHz data are at most the size of the symbols.

The flux density at 843 MHz from electrons accelerated continuously from a single source region in these same two cases – the freely-expanding wind and the confined wind – is shown in Figure 11. We have chosen $t_0$ to be 18.3 days in each case, and for the confined wind we use $r_w = 67 U_2 t_0$. Electron injection is assumed to begin after 913 days, roughly the time that the shock enters the stagnation zone (in the confined wind case), after which 1% of electrons crossing the shock are injected into the acceleration process. The electron diffusion coefficient is taken to be $\kappa_e = 10^{16}$ m$^2$ s$^{-1}$. For the case where the shock expands into a freely-expanding stellar wind, the flux density rises and then slowly decays, for essentially the reasons discussed after equation (17). The CR effects modify the details of the light curve, but not the qualitative trends. However, for acceleration at a shock expanding into a confined wind, the flux density continues to rise steadily over a very long timescale. This reflects the fact that both the magnetic field at the subshock, and the rate of electron injection (assuming injection of a constant fraction of the electrons encountered), decrease more slowly after the precursor encounters the termination shock than in the freely-expanding wind case, and eventually increase with time. Thus the effect of increased emission from freshly-injected electrons continues to dominate over the decreasing emission from those electrons downstream of the shock undergoing adiabatic losses.

4. Comparison with observations

To compare the theory with observations we calculate the electron acceleration and subsequent synchrotron emission as described in section 3, using the two-fluid results of section 2, making now specific assumptions about when and at what rates electrons are injected into the acceleration process. Three pieces of observational evidence suggest that the continually brightening radio emission from SNR 1987A comes not from a single source in which the rate of electron injection is steadily increasing, but rather that injection in discrete components or clumps has switched on at different times. Firstly, soon after the remnant was detected, the total radio flux density was observed to fluctuate on timescales of several days (Staveley-Smith et al. 1992), indicating that the emitting region was smaller than a spherically-symmetric shell of radius equal to that of the shock. Secondly, recent high-resolution images of the remnant show that the emission itself is clumpy (Staveley-Smith et al. 1993).

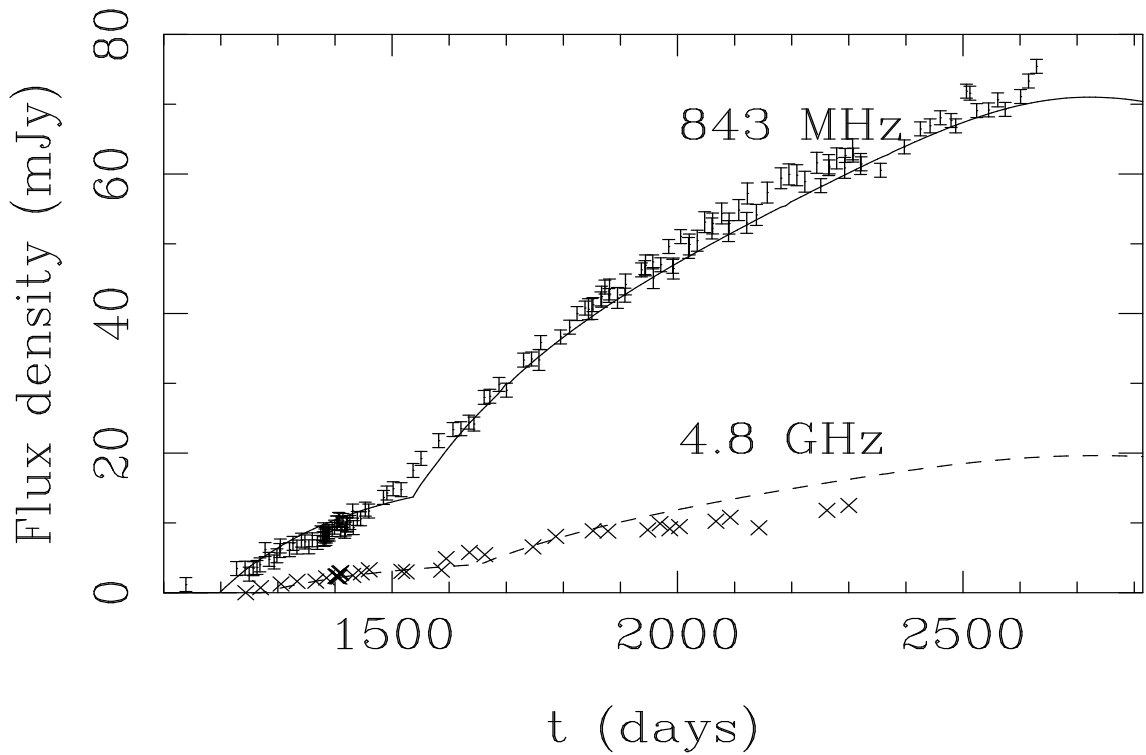

Fig. 13.— Best model fit to the observed radio emission from SNR 1987A, for acceleration at a CR-modified shock propagating into a confined wind.

Finally there has been at least one marked steepening of the light curve (around day 1450) when the rate of increase of the flux density jumped rapidly and then stayed at its new value, indicating a sudden change of injection (Staveley-Smith et al. 1992). We therefore model the radio source as two separate components in which electron injection starts at different times – perhaps on different parts of the shock – and which (some time later) produce the switch on or the break in the light curves when they commence emitting radiation at the observed frequency.

The value of the electron diffusion coefficient $\kappa_e$ is constrained by the observed delay in the switch on of emission at high frequency (4.8 GHz) compared to low frequency (843 MHz). Emission at 4.8 GHz first became detectable some 30–60 days after the switch on of emission at 843 MHz. ¿From equation (7) of BK, the ratio of the highest emitted frequencies at two successive times $t_1$ and $t_2$ (for acceleration at a shock of constant compression ratio) is given by

$$\frac{\nu_{\max}(t_1)}{\nu_{\max}(t_2)} = \frac{t_2}{t_1} \exp\left[2(t_1 - t_2)\Delta/t_c\right] \qquad (18)$$

where $t_c/\Delta$ is essentially the electron acceleration timescale. The observed delay implies that $t_c/\Delta = 34 - 68$ days. The observed frequency spectra suggest a value of $\rho = 2.7$ for the subshock compression ratio, and the two-fluid hydrodynamics solutions of section 2 give average values of $v_1 \sim 10^6 \, \mathrm{m\,s^{-1}}$ for the upstream fluid speed around the time that synchrotron emission was first observed. Together with the acceleration time, these values imply a range for the electron diffusion coefficient of $\kappa_e \sim 1.6 \times 10^{17} - 3 \times 10^{17} \, \mathrm{m^2\,s^{-1}}$. The actual switch on times themselves (as opposed to the delay between them) can then be fitted if electron injection begins somewhere between 2 years and 2.8 years after the explosion – the larger $\kappa_e$ the earlier the start of injection. The values of $\kappa_e$ derived here are smaller than those quoted in BK because the CR modifications imply a lower shock speed for a given ejecta speed. Furthermore, the shock continues to weaken and the fluid speeds continue to evolve after the switch on of radio emission. Therefore, since we assume that the electron diffusion coefficient doesn't change with time, $\kappa_e$ must be somewhat smaller than the above estimates in order to maintain the required average electron acceleration time.

The 'best' fit of the model flux from acceleration at a CR-modified supernova shock propagating out through a freely-expanding stellar wind is shown in Figure 12. The hydrodynamic solution used to produce the model light curves in Figure 12 is exactly that presented in Figure 2, with $t_0 = 18.3$ days. The electron parameters used are $\kappa_e = 10^{16} \, \mathrm{m^2\,s^{-1}}$ for both components; $t_{a1} = 2.5$ years and $\eta_{e1} = 0.008$ for the first component; and $t_{a2}$ = 2.75 years and $\eta_{e2} = 0.05$ for the second. The 843 MHz data plotted in Figure 12 (and later figures) are from Ball et al. (1994), while the data at 4.8 GHz are from Staveley-Smith et al. (1994).

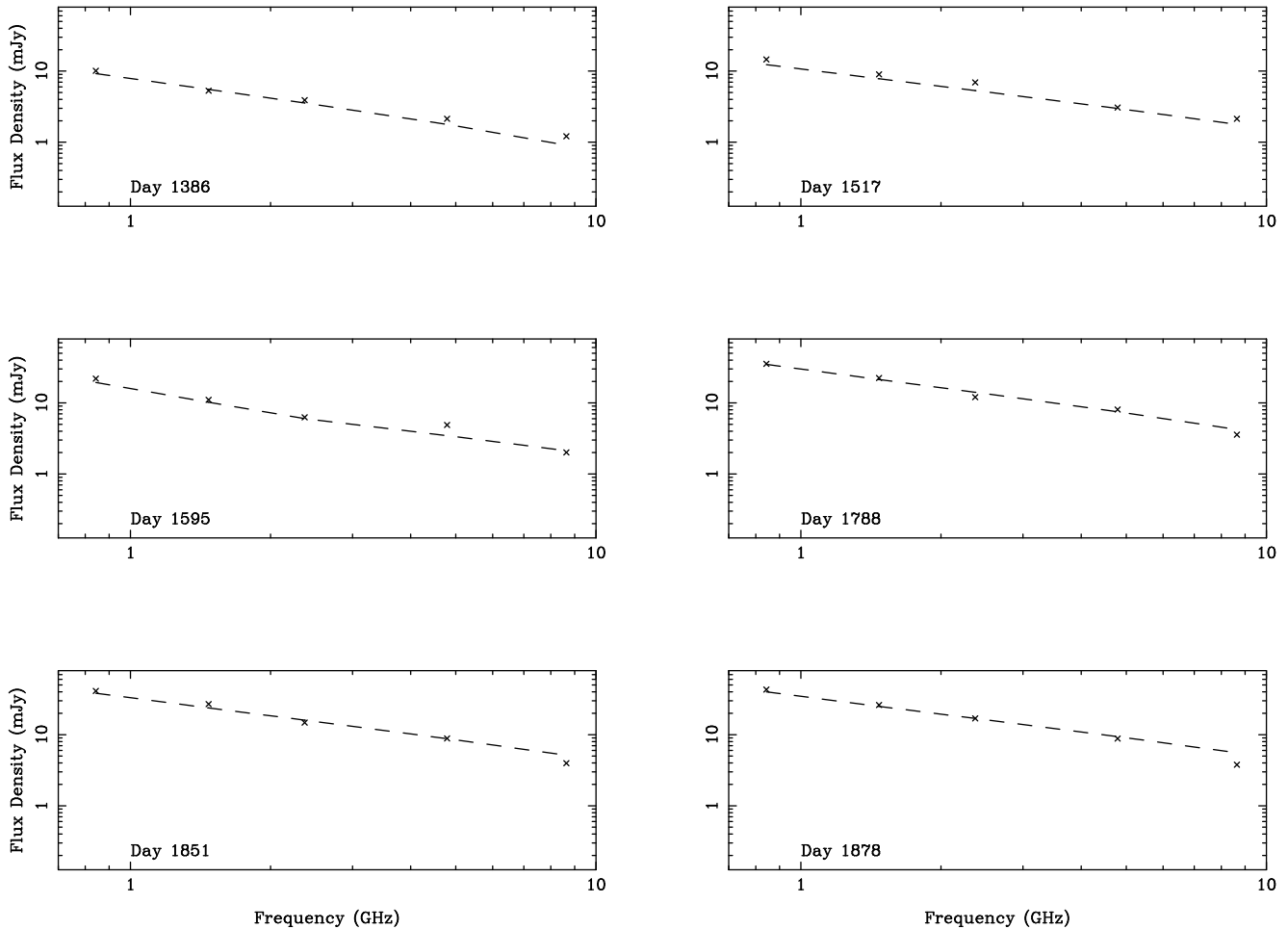

Fig. 14.— Frequency spectra from the 'confined wind' model (lines) versus the observed flux densities (points) at the three times for which observed spectra were presented by Staveley-Smith et al. (1992), and at three later times. The uncertainties in the data are smaller than the symbols used in the plots.

In this 'freely-expanding wind model' the emission from the first source component provides a good fit to the data up to the sudden jump in the rate of increase of the flux density at around day 1450–1500. Then with the addition of a second source component with the same value of $\kappa_e$, an acceptable fit is obtained up until about day 1700. This essentially reproduces the model fit of BK, and indicates that the CR modifications of the expanding shock can explain the relatively steep observed spectrum of the radio emission from SNR 1987A. However, for the reasons discussed in Section 3, the model light curves flatten off and eventually begin to decrease because of the decrease in the rate of injection of electrons, and in the magnetic field at the subshock. The observed flux however, has continued to increase steadily throughout the period of observations to date. We conclude from this that electron acceleration at a shock expanding into a freely-expanding wind is unlikely to be able to account for the continued increase in the radio emission from SNR 1987A. Of course, if, for some reason, the injection of electrons increases steadily, despite the decreasing flux of electrons across the subshock, then it would be possible to fit the observations. This case is, however, rather artificial – a corresponding continual increase in the proton injection rate would lead to a departure of the spectrum from that observed.

Figure 13 shows the 'best' fit of the model flux from acceleration at a CR-modified supernova shock propagating through a confined wind. The hydrodynamic solution used to produce the model light curves in Figure 13 is exactly that presented in Figure 5 with $t_0 = 18.3$ days and $r_w = 67 U_2 t_0$. Injection of electrons from the two source components in this model begins after 2.55 years and 2.9 years. The fraction of the electrons crossing the subshock that are injected are $\eta_{e1} = 0.006$ and $\eta_{e2} = 0.02$ respectively. The value of $\kappa_e$ is the same in the two source components, and is $\kappa_e = 10^{16}\,\mathrm{m^2\,s^{-1}}$. This 'confined wind' model provides an excellent fit to the observed radio emission, clearly reproducing the continuing increase of the flux density. The implied radius of the termination shock is $2 \times 10^{15}$ m. In Figure 13 we plot the predicted flux up to day 2900, about the time the shock reaches a radius of $6.3 \times 10^{15}$ m, which corresponds to the radius of the ring.

Figure 14 shows the correspondence between the frequency spectra obtained from the 'confined wind' model, and the observed spectra, at six different epochs. The data at 1.5 GHz and 2.4 GHz at the three later epochs are from Staveley-Smith (1994). By virtue of the reduced compression ratio of the CR-modified subshock, the model reproduces and provides a physical explanation for the relatively steep observed frequency spectra. While the absolute level of the radio emission is sensitive to the specific rates and timing of the electron injection, the synchrotron spectrum is essentially determined by the hydrodynamical solution for the CR-modified shock.

## 5. Discussion

The model for the radio emission from SNR 1987A due to BK, involving synchrotron emission from electrons accelerated at the expanding supernova shock, provides a good fit to the observations up to about day 1800. However it leaves unanswered the question of why the synchrotron spectrum is steeper than $\nu^{-0.5}$ as expected if the electron acceleration occurs at strong shock. In this paper we show that the inclusion of CR effects on the shock structure, via a two-fluid treatment, suggests that an initially strong SN shock can be substantially modified by the pressure exerted on the plasma upstream of the shock by the CRs which are also accelerated by the shock. This process can lead to a relatively rapid evolution of the shock. The most important effect for the electron acceleration is the formation of a substantial CR precursor with a width which is comparable to the shock radius. Our estimate of the mean free path for the radiating electrons is substantially smaller than that of the CRs. The electrons therefore see the precursor as a gradual change in the fluid properties and undergo diffusive acceleration only at the gas subshock. This has a much smaller compression ratio than that of an unmodified shock, and that of the shock-precursor system as a whole, providing a natural explanation for the apparent weakness of the shock responsible for the electron acceleration in SNR 1987A.

In order for CR modifications of the SN shock to account for the observed spectrum, the efficiency of CR injection needs to be sufficiently high for a precursor of thickness comparable to the shock radius to be established. If the levels of proton injection are too low or the value of $\kappa_p$ too high, there is essentially no modification of the shock. However, the levels of proton injection required to modify the shock sufficiently to account for the observed synchrotron spectrum are consistent with the levels assumed for CR production at supernova remnants in the Sedov-Taylor or adiabatic phase of evolution, which are generally believed to be the main source of galactic CRs at energies up to $10^{15}$ eV.

The simplest generalisation of the BK model, including CR modification of an initially strong shock but retaining the assumption that the shock propagates through a freely-expanding stellar wind from the progenitor star, leads to difficulties in explaining the continuing increase in the observed flux densities from SNR 1987A. In the original model the flux density at a given frequency peaks and then starts to decrease because of the decreasing magnetic field at the subshock. When CR modification effects are included the decrease occurs sooner, because the CR precursor accelerates the plasma ahead of the subshock, decreasing the flux of electrons encountering the shock. This causes the rate at which electrons are injected into the acceleration process to decrease with time, assuming injection of a fixed fraction of the electrons encountering the shock. The freely-expanding wind model (presented in Figure 12) therefore starts to lag significantly below the observed flux densities at times after about day 1700.

However, the wind from the blue giant progenitor of SN1987A may well be confined by an envelope of more dense material emitted during an earlier red giant phase. The expanding supernova shock may therefore encounter a termination shock in the wind, after which the density of the undisturbed circumstellar material is independent of radius and the magnetic field dependence changes from an $r^{-1}$ decrease to an increase as $r$. A model in which the electron acceleration occurs at a SN shock expanding through such a confined wind, with parameters appropriate to SNR 1987A, provides a very good fit to the observed radio emission at all frequencies. As in the original model of BK, two source components are needed to account for the sudden jump in the rate of increase of the flux density observed around day 1450–1500. However, an acceptable fit is obtained using the same value for the electron diffusion coefficient $\kappa_e$ in the the two source components, and with models in which the times at which electron injection in the two components are within 10% of each other, and are close to the time the subshock encounters the wind termination shock. The first point is important since it suggests that the physical conditions in the two source components need not be very different in order for this model to explain the observations. The latter feature implies that the start of electron injection in the two source components could occur at the same radius, with the apparent time difference in our observer's frame being due to light travel time effects. Furthermore, it suggests that the start of electron injection may be associated with the "impact" of the expanding SN shock on the wind termination shock. This encounter is a relatively unspectacular event, even somewhat poorly-defined, because the CR precursor of the SN shock smooths out the discontinuity in the circumstellar material before it is encountered by the subshock itself. Nevertheless, it is tempting to retain the original idea that the start of electron acceleration may be the result of the change in the conditions at the subshock when it encounters the termination shock of the wind (Ball & Kirk 1992b, Chevalier 1992a).

The model fits presented here do not unambiguously determine the position of the wind termination shock $r_{\rm w}$, but they do enable us to place limits on it. One difficulty which cannot be removed is that of the dependence of the results on $t_0$, the time when CR acceleration begins. However, it is clear that if the stagnation zone begins too close to the progenitor, the CR effects will completely smooth out the subshock, and hence electron acceleration will cease, on a timescale which is so short that it is in conflict with the observations of the continuing increase in the radio emission. On the other hand, if the wind termination shock is too far from the progenitor, the electron acceleration at the subshock continues but the decreasing magnetic field and electron injection again imply that the flux density peaks and starts to decline on timescales which are in conflict with the continuing increase observed. We therefore conclude that the wind termination shock is situated at a radius of about $2 \times 10^{15}$ m, roughly one third of the radius of the dense ring of material around SN1987A.

Our model for the radio emission suggests that the CR pressure behind the shock, i.e., in the downstream plasma, is quite high. When the shock encounters the relatively dense ring of material surrounding SN1987A, gamma-rays may be produced by the decay of pions resulting from the collision of CR protons with the dense population of protons in the ring. The CR pressure is of the same order as the gas ram pressure (Figure 1) so $P_C \sim \rho_1 v_1^2$. The proton number densities of the upstream gas in the precursor and in the dense ring are $\sim 16\,{\rm cm}^{-3}$ and $\sim 2 \times 10^4\,{\rm cm}^{-3}$ respectively (Fransson et al. 1989). The rate of gamma-ray emission above 1 TeV can then be estimated from Drury, Aharonian and Völk (1994) to be

$$\dot{N}(>1\,{\rm TeV}) = q_\gamma n E_C V \sim 3 \times 10^{34}\,{\rm s}^{-1}\;, \qquad (19)$$

where $q_\gamma$ is a constant, taken to be $10^{-17}$, containing information on the nuclear collision cross sections and shape of the CR spectrum and $n$ is the number density of protons in the ring which fills a volume $V$. The resulting flux at Earth is $9 \times 10^{-14}\,{\rm cm}^{-2}\,{\rm s}^{-1}$, somewhat below the threshold of current atmospheric Cerenkov detectors. The flux from the confined wind can easily be calculated and is slightly lower than this. Ideally, a more detailed analysis in which our simulations are continued until the CR modified precursor and subshock hit the dense ring, is required to check this estimate. However this is beyond the scope of the present paper. It is interesting to note that a similar estimate for SN 1993J results in a predicted TeV flux which may be detectable (Kirk, Duffy & Ball 1995).

The predictive power of our model is limited by the known departures from spherical symmetry of the distribution of circumstellar material as the SN shock approaches the ring and bipolar nebula. Furthermore, we expect the behaviour of the system to change character as the SN shock approaches the ring simply because the density of the circumstellar material encountered by the shock will then begin to increase rapidly. Our models suggest that the shock radius will be comparable to the ring radius some 8 years after explosion, i.e., in 1995. For the hydrodynamic solution of figure 5 in the confined wind case the subshock compression ratio will have decreased to a value of about 2 at this point. This would imply an eventual steepening, though not a disappearance, of the radio emission before the point of impact between the shock and the circumstellar ring.

## ACKNOWLEDGEMENTS

The authors thank Lister Staveley-Smith, Duncan Campbell-Wilson and David Crawford for providing data prior to publication, and for many helpful discussions. This collaboration was made possible by visits of JGK to Sydney supported by the Research Centre for Theoretical Astrophysics and of LB to Heidelberg supported by the Max Planck Institut für Kernphysik.

## 6. Appendix

### 6.1. Appendix A

All quantities in equations (2) to (7) can be written in terms of their value immediately upstream of the shock at $t_0$ (the time at which CR injection begins) times a function of $\tau \equiv t/t_0$. For example consider equation (2) and define $\hat{\rho} \equiv \rho/\rho_1(t_0)$. With $\hat{r}_s \equiv r_s/r_s(t_0)$ and $d\eta \equiv d\tau/\hat{r}_s$ the mass conservation equation becomes

$$\frac{\partial \hat{\rho}}{\partial \eta} + (\hat{U} - \xi \hat{U}_s)\frac{\partial \hat{\rho}}{\partial \xi} = -\frac{\hat{\rho}}{\xi^2}\frac{\partial}{\partial \xi}(\xi^2 \hat{U}) \quad (20)$$

---

This 2-column preprint was prepared with the AAS LATEX macros v3.0.

where $\hat{U} = U/U(t_0)$. Likewise all other equations governing the hydrodynamics can be rescaled to include only functions of $\tau$.

Equations (1) to (3) describing the dynamics of the massive, thermal component of the two fluid system are solved explicitly by finite differencing. The precursor region is divided into $M$ cells each of width $\Delta \xi_j$. If the solution is known at the $n$th timestep then the gas dynamic equations may be differenced to give the solution a time $\Delta t$ later at the $(n+1)$th timestep later at the $(n+1)$th timestep

$$\rho_j^{n+1} = \rho_j^n + \frac{\chi_j^n}{r_s}\frac{\Delta t^n}{\Delta \xi_{j+1/2}}\left(\rho_{j+1}^n - \rho_j^n\right)$$
$$-\frac{\rho_j^n}{r_s \xi_j^2}\frac{\Delta t^n}{\Delta \xi_j}\left(\xi_{j+1/2}^2 U_{j+1/2}^{n+1/2} - \xi_{j-1/2}^2 U_{j-1/2}^{n+1/2}\right) \quad (21)$$

$$U_j^{n+1} = U_j^n + \frac{\chi_j^n}{r_s}\frac{\Delta t^n}{\Delta \xi_{j+1/2}}\left(U_{j+1}^n - U_j^n\right)$$
$$-\frac{1}{r_s \rho_j^n}\frac{\Delta t^n}{\Delta \xi_j}\left(P_{Gj+1/2}^{n+1/2} - P_{Gj-1/2}^{n+1/2}\right) \quad (22)$$
$$-\frac{1}{2r_s \rho_j^n}\frac{\Delta t^n}{\Delta \xi_j}\left(P_{Cj+1}^n - P_{Cj-1}^n\right)$$

$$S_j^{n+1} = S_j^n + \frac{\chi_j^n}{r_s}\frac{\Delta t^n}{\Delta \xi_{j+1/2}}\left(S_{j+1}^n - S_j^n\right) \quad (23)$$

where $\chi_j^n = \xi_j U_s^n - U_j^n$ and the subscript $j$ refers to the $j$th spatial cell. The quantities $U_{j+1/2}^{n+1/2}$ and $P_{Gj+1/2}^{n+1/2}$ are the solutions of the Riemann problem between the $j$th and $(j+1)$th cells. Since this scheme is explicit it is subject to a Courant Friedrich Lewy (CFL) stability limit on the timestep

$$\Delta t_j^n < \frac{\Delta \xi_1}{\max(U_j^n + C_{sj}^n)} \quad (24)$$

with $C_{sj}^n$ the local sound speed. Critical to the numerical accuracy of our method is the resolution of the prescursor. Typically we use about 100 cells per precursor lengthscale which is sufficient to obtain accurate results. If we were to use an explicit scheme for the CR equation (4) the diffusion term would put a strict CFL condition on $\Delta t$ that would then scale with $(\Delta \xi_1)^2$. Therefore we use a Crank-Nicholson scheme where the CR spatial gradient is averaged over the timestep. For example the diffusion term in equation 4 is differenced to

$$\frac{1}{2r_s^{(n+1)2}\xi_j^2 \Delta \xi_j}\left[\frac{\xi_{j+1/2}^2 \overline{\kappa}_{j+1/2}^{n+1}}{\Delta \xi_{j+1/2}}\left(P_{Cj+1}^{n+1} - P_{Cj}^{n+1}\right)\right.$$
$$\left.-\frac{\xi_{j-1/2}^2 \overline{\kappa}_{j-1/2}^{n+1}}{\Delta \xi_{j-1/2}}\left(P_{Cj}^{n+1} - P_{Cj-1}^{n+1}\right)\right]$$
$$+\frac{1}{2r_s^{n2}\xi_j^2 \Delta \xi_j}\left[\frac{\xi_{j+1/2}^2 \overline{\kappa}_{j+1/2}^n}{\Delta \xi_{j+1/2}}\left(P_{Cj+1}^n - P_{Cj}^n\right)\right.$$

$$-\frac{\xi_{j-1/2}^2 \overline{\kappa}_{j-1/2}^n}{\Delta \xi_{j-1/2}} \left(P_{Cj}^n - P_{Cj-1}^n\right)\right] \quad (25)$$

With all other spatial gradients differenced in a similar manner the resulting numerical equation can be solved by tridiagonal matrix inversion and provides no additional stability constraint on the size of the timestep.

## 6.2. Appendix B

The characteristic equation of the acceleration equation (10) is $dp/dt = p\Delta/t_c$. Then, for constant $\kappa_e$ and $Q(t < t_a) = 0$, the solution to equation (10) can be written in the form

$$\begin{aligned}
N(t,\xi) &= \exp\left(-\int_0^t dt'\, j(\xi,t')\right) \times \\
&\quad \left\{\int_{t_a}^t dt'\, Q(t') \exp\left(\int_0^{t'} dt''\, j(\xi,t'')\right) \times \right. \\
&\quad \left. \delta\left[\xi \exp\left(\int_0^{t'} dt''\, g(\xi,t'')\right) - p_0\right]\right\} (26)
\end{aligned}$$

where $g(p,t) = \Delta/t_c$, $h(p,t) = P_{\rm esc}/t_c$, $j(p,t) = g(p,t) + p\,\partial g(p,t)/\partial p + h(p,t)$ and

$$\xi = p/\exp\left(\int_0^t dt'\, g(p,t')\right). \quad (27)$$

The solution of equation (13) proceeds as follows. If we express all functions in terms of the variables $p, \mathbf{R}, t$ instead of $p, \mathbf{r}, t$ then

$$\left(\frac{\partial}{\partial t}\right)_{\mathbf{r}} + \mathbf{v} \cdot \nabla_{\mathbf{r}} = \left(\frac{\partial}{\partial t}\right)_{\mathbf{R}} \quad (28)$$

and defining $D(\mathbf{R},t) = -\frac{1}{3}\nabla \cdot \mathbf{v}$ we can write equation (13) as

$$\left(\frac{\partial f(p,\mathbf{R},t)}{\partial t}\right)_{\mathbf{R}} + D(\mathbf{R},t)\, p\, \frac{\partial f(p,\mathbf{R},t)}{\partial p} = 0 \quad (29)$$

The method of characteristics then gives

$$\ln(p) = \int_0^t dt'\, D(\mathbf{R},t') + \text{constant}, \quad (30)$$

so introducing

$$\xi = p/\exp\left(\int_0^t dt'\, D(\mathbf{R},t')\right) \quad (31)$$

it follows that equation (29) can be written in the form $(\partial f/\partial t)_\xi = 0$, which has as its solution an arbitrary function of $\mathbf{R}$ and $\xi(p,\mathbf{R},t)$. The solution of interest follows from the boundary condition which relates $f[p, \mathbf{R}(r = r_s(t), t), t]$, where $r_s(t)$ is the radius of the shock, to $N(p,t)$.

As a specific example we consider the situation discussed by BK. where the downstream fluid speed is radial and independent of both $r$ and $t$, i.e. $\mathbf{v} = U_2\,\hat{\mathbf{r}}$. Equation (14) then reduces to $R = r - U_2 t$, and thus $D(R,t) = -2U_2/[3(R + U_2 t)]$, which on substitution into equation (31) gives $\xi = p[(R + U_2 t)/R]^{2/3}$. The downstream fluid speed $U_2$ is related to the shock speed $U_s$ by the relation $U_2 = U_s(\rho - 1)/\rho$. Therefore if the shock compression ratio is constant it follows that the shock speed $U_s$ is also constant and thus $r_s = U_s t$. The boundary condition is then

$$f[p, R(r = U_s t, t), t] = f_s(p,t). \quad (32)$$

The specific solution we require follows from equation (12) if we replace $p$ by the combination $p[(R + U_2 t)/R]^{2/3} F(R)$ where $F(R)$ is chosen such that $\{p[(R + U_2 t)/R]^{2/3} F(R)\}_{t=R/(U_s-U_2)} = p$. Thus $F(R) = \rho^{-2/3}$ and $f(p,R,t)$ is given by equation (15).